\documentclass[10pt]{iopart}

\usepackage{graphicx}
\usepackage{epsf}
\usepackage{xcolor}
\usepackage{bm}

\usepackage[utf8]{inputenc}
\usepackage[T1]{fontenc}
\usepackage{mathptmx}

\usepackage{booktabs,caption}
\usepackage[flushleft]{threeparttable}

\usepackage{amssymb}

\newcommand{\thbn}{\theta_{\rm Bn}}

\newcommand{\omci}{\Omega_\mathrm{i}}

\newcommand{\mss}{M_\mathrm{s}}

\newcommand{\ma}{M_\mathrm{A}}
\newcommand{\mi}{m_\mathrm{i}}
\newcommand{\me}{m_\mathrm{e}}
\newcommand{\mpp}{m_\mathrm{p}}
\newcommand{\lse}{\lambda_\mathrm{se}}

\newcommand{\vsh}{v_\mathrm{sh}}

\newcommand{\ti}{T_\mathrm{i}}
\newcommand{\te}{T_\mathrm{e}}

\newcommand{\mnras}{MNRAS}
\newcommand{\apjl}{ApJL}
\newcommand{\apj}{ApJ}
\newcommand{\prl}{Phys. Rev. Lett.} 

\newcommand{\pjm}{\textcolor{black}}

\newcommand{\rev}{\textcolor{black}}

\begin{document}

\title[Electron acceleration at supernova remnants]{Electron acceleration at supernova remnants}

\author{Artem Bohdan}

\address{Deutsches Elektronen-Synchrotron DESY, Platanenallee 6, 15738 Zeuthen, Germany}
\ead{artem.bohdan@desy.de, artem.bohdan@ipp.mpg.de}
\vspace{10pt}
\begin{indented}
\item[] Accepted for publication in Plasma Physics and Controlled Fusion
\end{indented}

\begin{abstract}
Supernova remnants (SNRs) are believed to produce the majority of galactic cosmic rays (CRs). SNRs harbor non-relativistic collisionless shocks responsible for acceleration of CRs via diffusive shock acceleration (DSA), in which particles gain their energies via repeated interactions with the shock front. As the DSA theory involves pre-existing mildly energetic particles, a means of pre-acceleration is required, especially for electrons. Electron injection remains one of the most troublesome and still unresolved issues and our physical understanding of it is essential to fully comprehend the physics of SNRs. To study any electron-scale phenomena responsible for pre-acceleration, we require a method capable of resolving these small kinetic scales and Particle-in-cell (PIC) simulations fulfill this criterion. Here I report on the latest achievements made by utilising kinetic simulations of non-relativistic high Mach number shocks. I discuss how the physics of SNR shocks depend on the shock parameters (e.g., the shock obliquity, Mach numbers, the ion-to-electron mass ratio) as well as processes responsible for the electron heating and acceleration.
\end{abstract}

%
%
%
%
\ioptwocol

\section{The electron injection problem}

More than a century ago \pjm{the} Hess balloon experiments \pjm{provided evidence that} proved the existence of ionized radiation at altitudes above 1 km \cite{Hess}, clearly indicating its extraterrestrial origin. It is now well established that the cause of these measurements is cosmic rays (CRs) composed of high energy particles reaching the Earth from space.  Despite the fact that CRs are studied for over a century, their origin remains one of \pjm{most fundamental} unsolved problems in modern astrophysics.

Low energy CRs are produced by the Sun, while CRs with energies above $10^9$ eV originate from one of the astrophysical accelerators, such as supernova remnants (SNRs), pulsar wind nebulae, jets of active galactic nuclei, gamma-ray bursts, galaxy clusters etc. Among these sources, SNRs are always of particular interest, because they are known as efficient CR accelerators and a source for the \pjm{majority} of galactic CRs due to nonthermal radiation in various wavebands \cite{Clark,Koyama,Aharonian}. Since the late 70s it has been known that CR particles are accelerated via diffusive shock acceleration (DSA), a first-order Fermi process (see, e.g., \cite{Axford,Blandford,Drury,Bell1,Bell2}), in which particles gain their energies \pjm{from repeated} interactions with the shock front resulting from a supernova explosion. It is crucial, however, that DSA works only for high-energy particles whose energy is larger than the injection energy ($\varepsilon_{inj}$), and whose gyroradius is larger than the width of the shock transition layer. Therefore, particles should be preaccelerated or, in other words, injected into the DSA process. This puzzling and still unresolved issue is known as the injection problem.

\begin{figure}
    \centering
    \includegraphics[width=.99\linewidth]{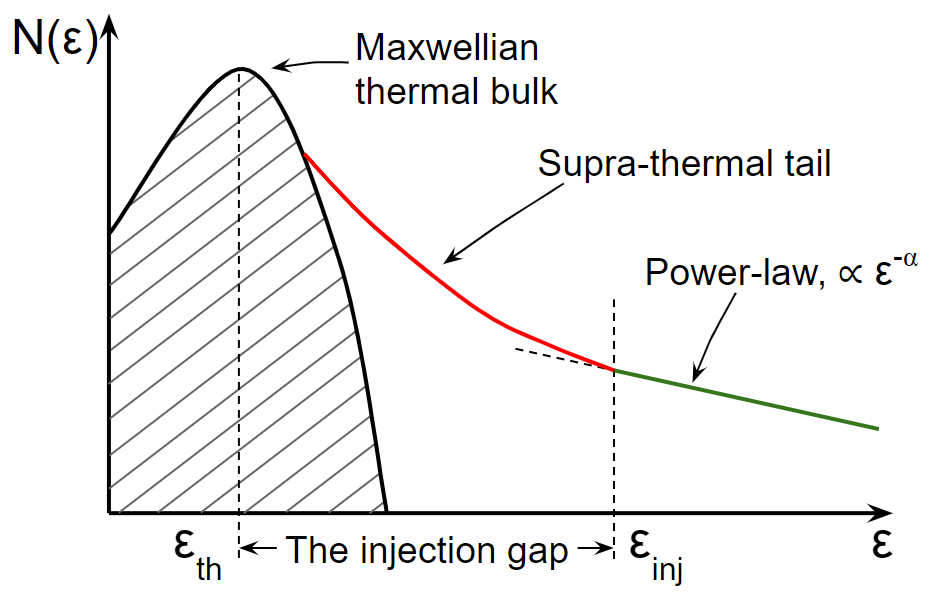}
    \caption{Particle energy spectrum at the shock with internal particle injection mechanisms \cite{Giacalone}. $\varepsilon_{inj}$ is the injection energy, $\varepsilon_{th}$ is the energy of the thermalized downstream plasma.}
    \label{inj_spectra}
\end{figure}

The particle energy distribution at a shock with an efficient particle acceleration mechanism consists of three particle populations \cite{Giacalone} (Fig.~\ref{inj_spectra}): heated at the shock thermal Maxwellian bulk, the supra-thermal tail of preaccelerated particles (red curve), and high-energy particles with energies above $\varepsilon_{inj}$ and \pjm{those} which are represented by a power-law distribution (green line). A particle during an injection process is picked up from the thermal bulk and accelerated up to the injection energy jumping over the injection gap. Since plasma consists of various particle species (e.g., protons, electrons) the injection processes are also different for them. For example, a proton is already injected if its gyroradius is a few times larger than the shock ramp width $r_{inj,p} \approx \xi d_{ramp}$, where $\xi \approx 3$. The shock ramp is the region where rapid changes of plasma density, temperature and velocity happen; its width approximately equals the gyroradius of the thermal proton $r_{th,p}$. Therefore, the proton injection energy is about 10 times larger than the thermal energy ($\varepsilon_{inj,p}/\varepsilon_{th,p} = \xi^2 \approx10$) and this injection gap can be overcome with one or two cycles of the shock drift acceleration (SDA \cite{Caprioli2015}). However, the situation with electrons is much more difficult. The gyroradius or momentum of the injected electron by definition is the same as for the proton ($p_{inj,e} = p_{inj,p}$), which implies a much higher injection energy for an electron and the electron injection gap is considerably larger.
The ratio of the initial electron energy to the electron injection energy depends on the proton-to-electron mass ratio and the shock speed and can be estimated as:
\begin{equation}
\frac{\varepsilon_{inj,e}}{\varepsilon_{th,e}} \approx 2\xi \frac{\mpp}{\me} \frac{c}{\vsh} \approx 10^5-10^6 \ . 
\end{equation}
Such a huge injection gap is the reason why the electron injection problem is extremely challenging.

For a proper modeling of the nonthermal radiation emitted by relativistic electrons we should know how many of them are involved in radiation processes. Therefore we should explain how electrons are accelerated through all these orders of magnitude in energy up to the injection energy where they continue to be accelerated via a much better understood DSA process. The injection problem can be solved \rev{by} revealing mechanisms responsible for electron energization which usually happens in two stages: thermalization (redistribution of the upstream electron energy and transfer of the proton upstream energy) and consequent acceleration via shock internal mechanisms.

\begin{figure*}[!t]
    \centering
    \includegraphics[width=.99\linewidth]{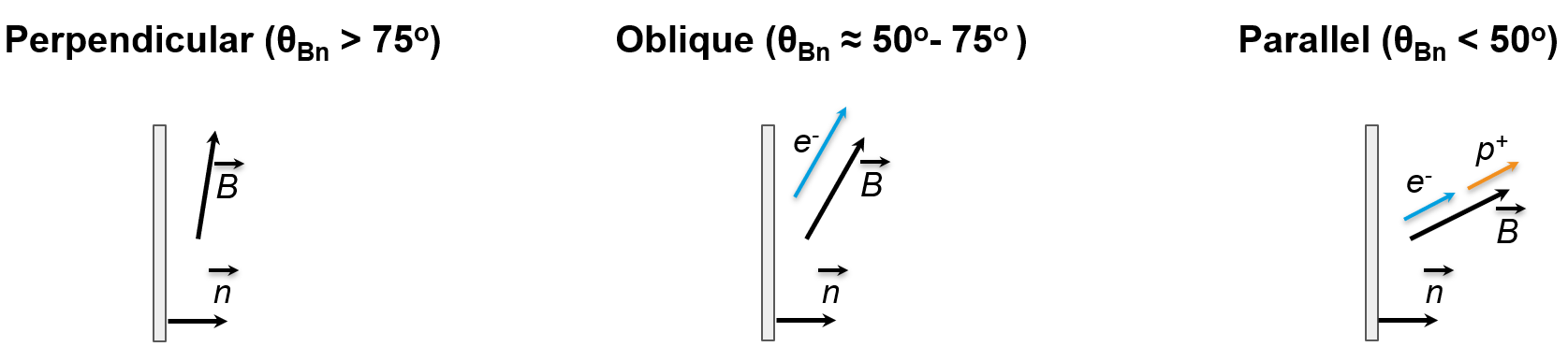}
    \caption{No particles can escape towards the shock upstream in perpendicular shocks, only electrons can escape in oblique shocks, both ions and electrons can escape in parallel shocks.}
    \label{figElectrostaticGamma}
\end{figure*}

The injection problem requires a tool capable of describing \pjm{the} entire shock microphysics accounting for all participating particle species. \rev{Observations \cite{vanAdelsberg,Ghavamian2013,Liu} and laboratory experiments \cite{Fiuza2020,Matsukiyo2022} combined with theoretical studies \cite{Fiuza2012,Korneev,Fox2017} can reveal some aspects of shock physics, however,} due to their limited resolution \rev{or restrictions of laboratory environment} can only approximately show where and how particles are accelerated. The only possible way to uncover the still hidden electron acceleration microphysics is plasma simulations, such as the fully-kinetic treatment, e.g., particle-in-cell (PIC) simulations \cite{Dawson,Birdsall}, which consider all particle species as individual particles moving in the self-generated electromagnetic field. PIC simulations allow us to obtain all necessary information about particles and electromagnetic fields at any given point in space and time and, therefore, to describe all details of particle acceleration processes. It makes this technique a core tool for the solution of the electron injection problem.

In this report we discuss how the electron injection problem is understood from the fully-kinetic plasma simulation point of view focusing on the structure of nonrelativistic SNR shocks and electron heating/acceleration processes operating there.

\section{SNR shock physics}

The interaction of supernova ejecta with the interstellar medium after supernova's explosions results in SNR shocks. It is well known that SNR shocks propagate with nonrelavitistic velocities \cite{Wang} ($\vsh \approx (1000-10000)$km/s~$= (0.003-0.03)c$, where $c$ is the speed of light) and are characterized by high sonic and Alfv\'enic Mach numbers ($M_s,M_A\approx20-1000$). 
High Mach number shocks are supercritical \cite{Marshall}, which means that the part of the upstream kinetic energy is dissipated via particle reflection by the shock potential hosting suitable conditions for various types of two-stream instabilities. Depending on the shock obliquity angle $\thbn$ (the angle between the upstream magnetic field and the shock normal vector) the shock reflected particles (protons, electrons, etc.) may escape the shock driving waves and disturbing the upstream medium the shock propagate through (e.g., \cite{Treumann}). Based on $\thbn$, shocks can be splitted into three classes: perpendicular (${75}^\circ  \lesssim  \thbn\le{90}^\circ$), oblique $(50^\circ\lesssim\thbn\lesssim{75}^\circ)$ and parallel ($0^\circ\le\thbn\lesssim{50}^\circ$) shocks. Note that the choice of boundaries among three shock classes is not very strict and may depend on the upstream conditions or the shock parameters.

Perpendicular shocks are characterised by a sharp shock transition for which the width is of the order of the upstream proton gyroradius ($d_{sh} \sim r_{p}$), because neither electrons nor protons can escape the shock towards the upstream. These shocks have been thoroughly studied using PIC simulation \cite{Amano2009,Kato2010,Matsumoto2012,Matsumoto2013,Matsumoto2015,Wieland2016,Bohdan2017,Bohdan2019a,Bohdan2019b,Bohdan2020a,Bohdan2020b,Bohdan2021} and results are described in Section~\ref{sec:perp}. The shock physics become much more difficult when $\thbn$ is small enough for energetic particles to escape a shock traveling far upstream, therefore the shock transition becomes much longer ($d_{sh}\gg r_{p}$). The narrow shock foot is replaced by a broad, turbulent foreshock (also called as the shock precursor) that extends far into the upstream flow. At oblique shocks only electrons escape the shock forming an extended electron foreshock filled with electrostatic and electromagnetic waves resulting in efficient electron scattering. These shock have been studied with PIC simulation only over the last five years \cite{Matsumoto2017,Xu2020,Kumar2021,Bohdan2022,Morris2022} and results are discussed in Section~\ref{sec:obl}. At parallel shocks the foreshock physics is dominated by ions and it was mostly studied with hybrid simulations \cite{Caprioli2014a,Caprioli2014b,Caprioli2014c} where ions are represented as particles and electrons are the massless fluid. As for PIC simulations, only a few papers \cite{Park2015,Kato2015,Arbutina2021a,Arbutina2021b} are dedicated for parallel shocks studies due to their extreme complexity and significantly higher computational costs compared to perpendicular shock simulations. Results \pjm{from these} are discussed in Section~\ref{sec:par}.

\section{Perpendicular shocks} \label{sec:perp}

\subsection{Shock structure}

\begin{figure*}
    \centering
    \includegraphics[width=.99\linewidth]{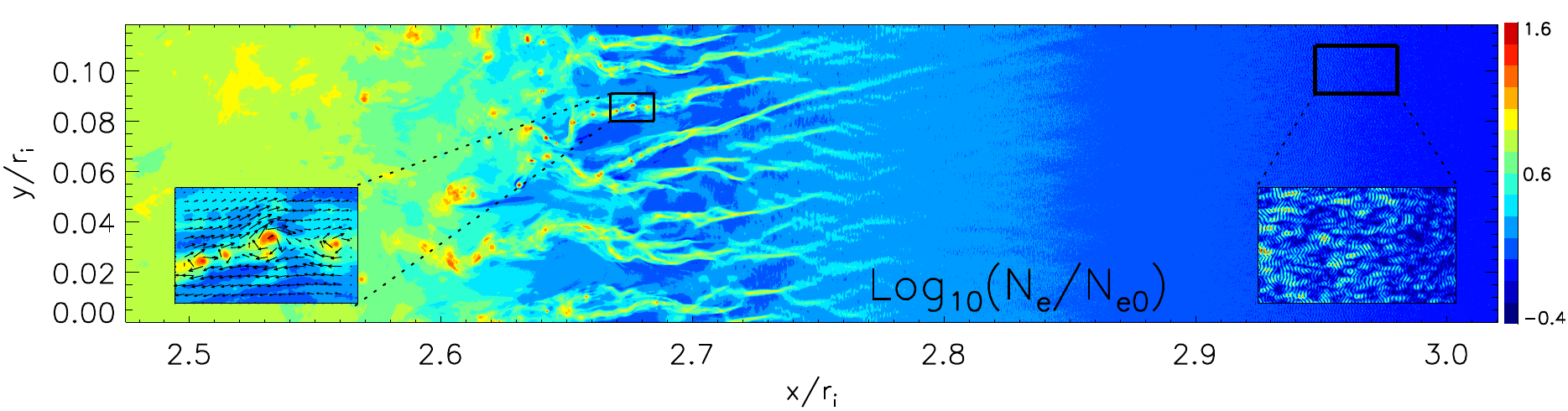}
    \caption{\emph{Perpendicular shock.} Main panel - electron density in the shock region; left inset panel - electron density and the in-plane component of magnetic field (arrows) in a region of the Weibel filament with magnetic reconnection; right inset panel - electrostatic field strength in the Buneman instability region. The shock transition includes the foot, ramp and overshoot regions and it spans from $x/r_i=2.5$ to $x/r_i=3$, the upstream is at $x/r_i>3$ and the downstream is at $x/r_i<2.5$. The shock parameters are $\vsh=0.263c$, $\ma=68.9$, $\mss=106$, $\mi/\me=400$ \cite{Bohdan2019a}.}
    \label{shock_str}
\end{figure*}

In perpendicular shocks the reflected ions can not escape the shock and after half a gyration in the upstream region they are advected downstream of the shock. On their way reflected ions interact with incoming ions and electrons driving two-stream instabilities. Perpendicular shocks are mediated by the electrostatic Buneman \cite{Buneman} and the ion-ion two-stream Weibel \cite{Weibel,Fried} instabilities making the shock a highly dynamical and complex system (Fig.~\ref{shock_str}). \rev{Magnetic field inside the shock is strongly amplified by Weibel instability, $B_{sh}/B_0 \propto \sqrt{\ma}$  \cite{Bohdan2021}, allowing for magnetic reconnection \cite{Matsumoto2015,Bohdan2020a} and efficient particle scattering.} \rev{It is interesting, that the growth rate of Weibel instability (if normalised to the ion upstream gyrotime) does not depend on the unrealistically high shock velocity or the reduced ion-to-electron mass ratio which permits the direct comparison of PIC simulations with real shocks \cite{Bohdan2021}.}
The resulting shock width is about $d_{sh} \approx 0.5r_i$, where $r_i$ is the upstream ion gyroradius. Comparison of 2D \cite{Bohdan2017} and 3D \cite{Matsumoto2017} simulations have demonstrated that 2D PIC simulations can reproduce realistic 3D physics of these shocks and most of the 3D shock physics can be studied with much less computationally demanding 2D simulations. Note that simulation by Matsumoto is done with oblique configuration ($\thbn \approx 74.3^\circ$), however the simulation time is short and the shock structure is still representative for perpendicular shocks. The presented in Figure~\ref{shock_str} shock structure is also supported by laboratory experiments \cite{Fiuza2020} as well as in-situ measurements \cite{Sundberg,Bohdan2021} demonstrating that perpendicular high Mach number shocks are indeed Weibel mediated and the presented shock structure is most likely physically correct.

\begin{figure}
    \centering
    \includegraphics[width=.99\linewidth]{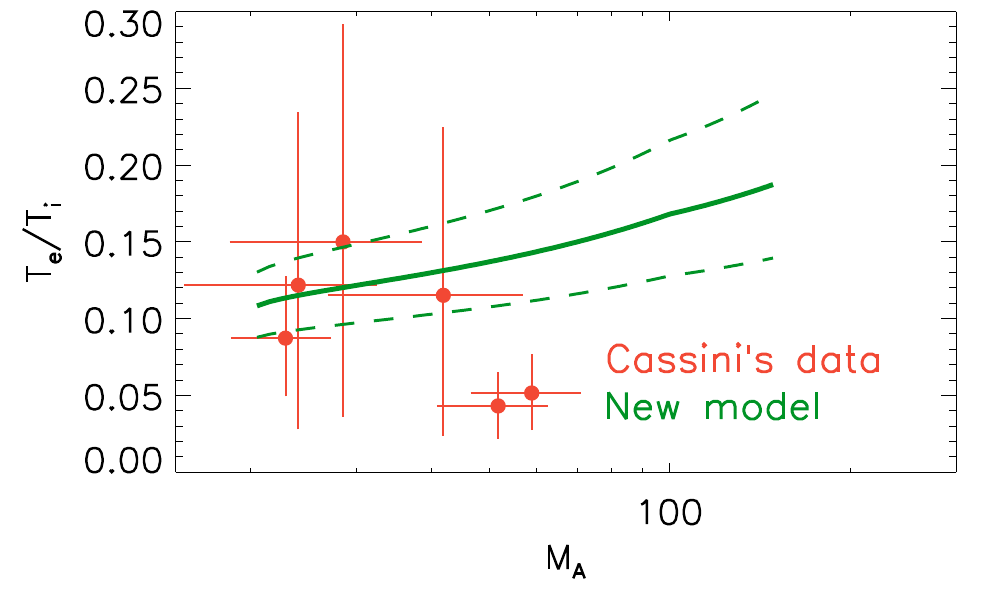}
    \caption{The electron heating model derived using PIC simulation of perpendicular shocks \cite{Bohdan2020a} (green) and Cassini's in-situ measurements \cite{Masters2011} (red). The green solid line represent the model, green dashed lines represent the one-sigma deviation levels to the model.}
    \label{heat_model}
\end{figure}

\subsection{Electron heating}

The shock converts a part of the upstream kinetic energy into the thermal particle energy. If there is no energy exchange between ions and electron inside the shock transition, the Rankine–Hugoniot jump conditions for temperature predict the downstream temperature ratio of $\te/\ti = \me/\mi$. However both in-situ measurements of the high Mach number Saturn's bow shock \cite{Masters2011} and telescope observations \cite{vanAdelsberg,Ghavamian2013,Liu} of SNR shocks show that some energy is transferred from ions to electrons and the resulting temperature ratio is larger than expected $\te/\ti\approx (0.05-0.5) \gg \me/\mi$.

\cite{Bohdan2020b} proposed an analytical model of electron heating in order to explain the temperature ratio measured using data taken by the Cassini spacecraft \rev{which measures plasma parameters on scales close to PIC simulations}. It shows that  the ion flow energy is transferred to electrons  via five channels: shock-surfing mechanism, the shock potential, magnetic reconnection, stochastic Fermi acceleration, adiabatic compression. This model accounts for dependencies of individual heating processes on the shock parameters, such as $\vsh$, $\ma$, $\mss$, $\mi/\me$, $\beta$. The new electron heating model predicts simulation results in a wide range of shock parameters, moreover, after \rev{using realistic shock velocity and the proton-to-electron mass ratio} it estimates the downstream temperature ratio in the range of $\te/\ti\approx (0.09-0.25)$ which matches Cassini's in-situ measurements within errorbars (Fig.~\ref{heat_model}).  Note, that predictions of the new model may deviate from some in-situ measurements \rev{(see two measurements with $\te/\ti\approx 0.05$ in Fig.~\ref{heat_model})} since PIC simulations consider shocks propagating through the ideal homogeneous medium and the shock structure should not deviate too much from the structure discussed above. \rev{Also, the Cassini spacecraft was able to measure reliably only the downstream electron temperature while other plasma and shock parameters are determined by the Saturn's bow shock model, which may introduce additional systematic errors explaining observed mismatch.}

\rev{Although $\te/\ti$ derived from telescope observations is similar to those from PIC simulation, it remains debatable how to execute a proper comparison since these approaches cover temporal and spatial scaled orders of magnitude apart from each other. It would require much longer PIC simulations and better modeling of electron behavior in the shock downstream including synchrotron cooling, charge exchange, Coulomb interaction, etc.}

\subsection{Electron Acceleration} \label{sec:perp_acc}

At perpendicular shocks electrons can be accelerated via the shock-surfing acceleration (SSA) mechanism at the Buneman wave region \cite{Matsumoto2012,Bohdan2017,Bohdan2019a}, magnetic reconnection \cite{Matsumoto2015,Bohdan2020a} and the stochastic Fermi-acceleration (SFA) mechanism \cite{Bohdan2017,Bohdan2019b}.

SSA is capable of accelerating electrons to energies tens of times larger than the upstream electron bulk energy for a wide range of shock Mach numbers if the so-called trapping condition is satisfied \cite{Matsumoto2012,Bohdan2019a}. However, in case a realistic mass ratio is applied SSA preacceleration can be heavily spoiled by the heating processes discussed above  \cite{Bohdan2019b}. 

During magnetic reconnection electrons can be accelerated via a number of mechanisms \cite{Matsumoto2015,Bohdan2020a}: interaction with outflows, X-points, first-order Fermi acceleration, vortex contraction, SFA. For higher Mach numbers magnetic reconnection becomes particularly active due to strong saturation of Weibel instability and almost all electrons participate in magnetic reconnection if $\ma \geq 70$ \cite{Bohdan2020a}. However, the overall acceleration efficiency via magnetic reconnection is going down due to a smaller amount of available magnetic field energy \cite{Bohdan2021} which potentially can be converted into electron heating and acceleration. 

The most efficient electron acceleration mechanism in perpendicular shocks is SFA. It is responsible for the production of the most energetic electrons; the maximal energy is proportional to the ion-to-electron mass ratio \cite{Bohdan2019b}. Therefore, even if the realistic mass ratio is applied, the downstream electron spectra will contain suprathermal electrons with energies much larger than those of thermal electrons.

The resulting suprathermal electron fraction is usually of the order of $0.5\%$ \cite{Bohdan2019b}. Most of them are accelerated via SFA especially in simulations with high mass ratios. However, the biggest problem with perpendicular shock\pjm{s} is that in this case no particles can escape the shock and the turbulence necessary for DSA can not be triggered self-consistently. Therefore, the maximal achieved energy of electrons is just 10-100 times larger than the thermal energy of the downstream electrons and the injection does not occur. To be noted that DSA can potentially be triggered in perpendicular shocks using additional ingredients such as seed CRs \cite{Caprioli2018}, neutral particles \cite{Ohira2016} or preexisting turbulence \cite{Trotta2021}.

\begin{figure*}[!t]
    \centering
    \includegraphics[width=.99\linewidth]{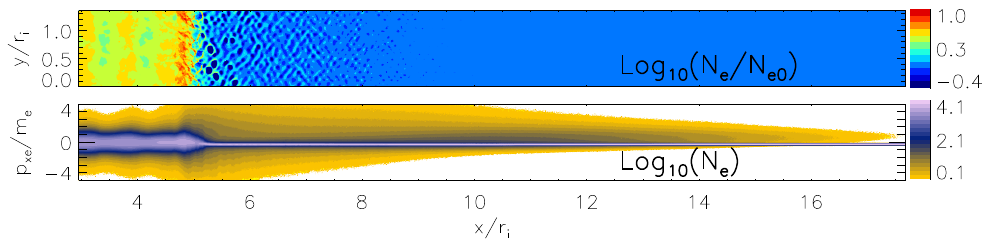}
    \caption{
    \emph{Oblique shock.} Top panel - electron density in the shock region; bottom panel - the $x$-$p_x$ phase-space density ($p_x$ is the particle momentum in x-direction) for electrons. The shock transition includes the electron foreshock, ramp and overshoot regions and it spans from $x/r_i\approx 4.5$ to $x/r_i\approx17$ , the upstream is at $x/r_i>17$ and the downstream is at $x/r_i<4.5$. The hock parameters are $\vsh=0.263c$, $\ma=30$, $\mss=32.5$, $\mi/\me=50$ \cite{Bohdan2022}.}
    \label{oblique_shock}
\end{figure*}

\section{Oblique shocks} \label{sec:obl}

\subsection{Shock structure}

As it was mentioned above, in oblique shocks fast electrons can escape the shock towards the upstream region (Fig~\ref{oblique_shock}, bottom panel). Therefore, the shock transition becomes much wider, and the shock foot is replaced by a broad, turbulent foreshock that extends far into the upstream flow generating electrostatic and electromagnetic waves. The shock width in this case is about $d_{sh} \approx 12.5r_i$. This structure captured both with 1D \cite{Xu2020,Kumar2021} and 2D \cite{Bohdan2022,Morris2022} PIC simulations. The only 3D simulation \cite{Matsumoto2017} is unfortunately too short to capture the electron foreshock structure. Note that around the shock transition (see Fig\ref{oblique_shock}, $x/\lse \approx (1000-1100)$) both Buneman and Weibel instability can be found, however their structure is somewhat modified due to waves excited in the foreshock.

In \cite{Bohdan2022} periodic boundary condition simulations were used to identify new instabilities appearing in the foreshock. The phase-space distribution of both particle species in the near and far foreshock regions are derived with the shock simulation and then used as initial conditions for simulations with periodic boundary conditions. We find that the observed electron-beam instabilities agree with the predictions of a linear dispersion analysis very well: the electrostatic electron-acoustic instability dominates in the far upstream of the foreshock, while the denser electron beams in the near upstream drive the oblique-whistler instability gyroresonant with the shock reflected electrons. \rev{Whistlers are responsible for creation of the foreshock turbulence with $\delta B/B \sim 1$. The magnetic field is likely amplified by a combination of whistler waves and Weibel instability.}
At the latest stages of the shock simulation the energy distribution of the escaping electrons becomes stable and allows for extrapolation further upstream. Knowing the triggering conditions for the whistler waves and the electron acoustic instability, one can estimate the simulation time needed to cover the entire foreshock in its steady state. Slightly more than $300 \omci^{-1}$ is required \cite{Bohdan2022} to cover the entire electron foreshock and reach the steady-state stage of the shock with $\ma = 30$. The shock width in the steady state is expected to be about $d_{sh} \approx 200r_i$. It poses one of the biggest challenges to study oblique shocks because their evolution can not be captured within several $\omci^{-1}$ like perpendicular shocks and they require substantially \pjm{more} computational resources.

\subsection{Electron heating}

\begin{figure}
    \centering
    \includegraphics[width=.99\linewidth]{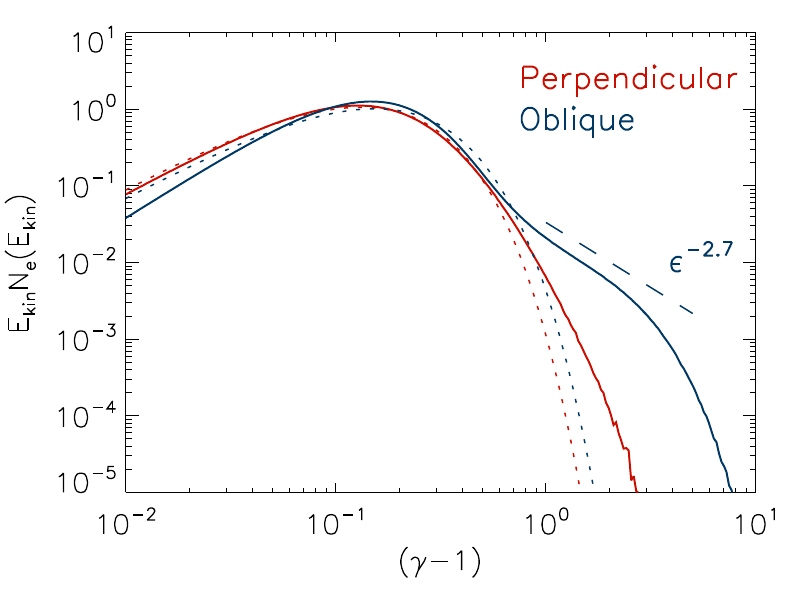}
    \caption{The downstream electron spectra for the perpendicular (red, $\thbn=90^\circ$) and oblique (blue, $\thbn=60^\circ$) shocks. The dotted lines represent fits of a relativistic Maxwellian to the low-energy part of the spectra. The straight dashed line denotes a power-law $\varepsilon^{-2.7}$.
    Perpendicular shock parameters are $\vsh=0.263c$, $\ma=22.6$, $\mss=35$, $\mi/\me=50$ \cite{Bohdan2019a}. Oblique shock parameters are $\vsh=0.263c$, $\ma=30$, $\mss=32.5$, $\mi/\me=50$ \cite{Bohdan2022}.}
    \label{obl_vs_perp_temp}
\end{figure}

So far there is no study dedicated to electron heating in oblique high Mach number shocks, however some indirect results can be discussed here. Firstly, the downstream electron temperature for 2D simulations of perpendicular shock \cite{Bohdan2019a} is consistent with that for the oblique shock from \cite{Bohdan2022} within 20\% margin (Fig.~\ref{obl_vs_perp_temp}). Note that these two simulations utilize almost the same set of plasma parameters.
Secondly, the data of Cassini spacecraft for Saturn's bow shocks shows that measurements for different obliquity angles \cite{Masters2011} are characterised with similar values consistent within errorbars. The expected electron-to-ion temperature ratio would be about $\te/\ti\approx (0.1-0.3)$. However, this topic requires extensive studies also because we still don't know how much energy \pjm{can be provided by the} different mechanisms \pjm{that} contribute to\pjm{ward} electron heating, especially processes connected to the recently identified whistler and electron-acoustic waves in the foreshock region.

\subsection{Electron Acceleration}

Since the structure of oblique shocks is more complicated compared to perpendicular shocks, oblique shocks provide a few more possible mechanisms for electron acceleration. In addition to already mentioned in Section~\ref{sec:perp_acc} SSA, SFA and magnetic reconnection, electrons can be energized via interactions with whistler and electron-acoustic waves, and stochastic shock drift acceleration (SSDA).

Interaction with the electron-acoustic mode may deflect the upstream electrons and sometimes even turn them back upstream contributing to the beam of shock reflected electrons \cite{Morris2022}. As a result, electrons may increase their energy by a factor of $\sim 10-50$. However, this mechanism does not contribute a lot (less than 0.1\% of incoming electrons are affected) because the electron-acoustic waves carry a little energy compared to the electron upstream kinetic energy.

Since whistler waves in the foreshock are gyroresonant with the shock reflected electrons, they should strongly affect electron trajectories providing efficient scattering and, possibly, DSA-like acceleration. As a result, the nonthermal electron population which can be represented with the power-law distribution with the index of about 2.7 (Fig.~\ref{obl_vs_perp_temp}), roughly  consistent with previous 1D simulations by \cite{Xu2020}, is formed. This acceleration mechanism can be particularly efficient in shocks with high Mach numbers accelerating up to 7\% of electrons.

\begin{figure*}[!t]
    \centering
    \includegraphics[width=.99\linewidth]{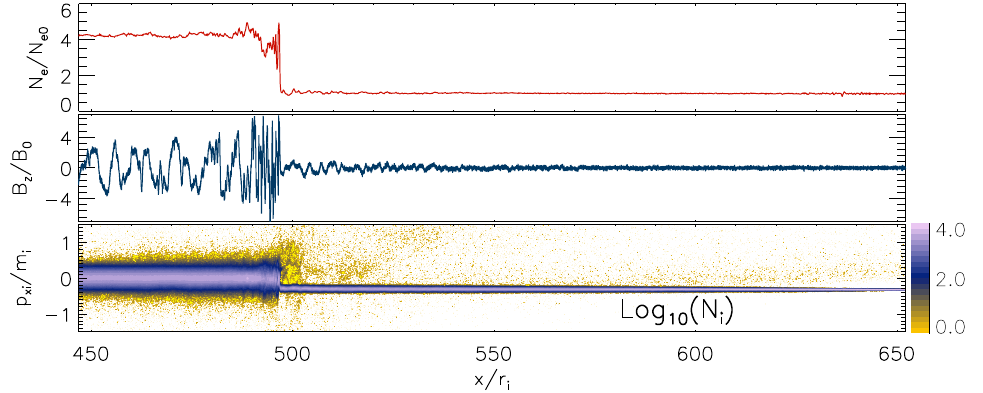}
    \caption{
    \emph{Parallel shock.} Top panel - electron density profile in the shock region; middle panel - $B_z$-profile in the shock region; bottom panel - the $x$-$p_{xi}/\mi$ phase-space density ($p_{xi}$ is the ion momentum in x-direction) for ions. The shock transition includes the ion foreshock, ramp and overshoot regions and it spans from $x/r_i\approx 490$ to $x/r_i\approx650$ , the upstream is at $x/r_i>650$ and the downstream is at $x/r_i<490$. The shock parameters are $\vsh=0.4c$, $\ma=16$, $\mss=35$, $\mi/\me=16$ \cite{Arbutina2021a,Arbutina2021b}.}
    \label{parallel_shock}
\end{figure*}

Another very promising candidate for electron injection in oblique shocks is SSDA, firstly observed in 3D shock simulation by \cite{Matsumoto2017}. It is a combination of the classical DSA with the pitch-angle scattering provided by waves excited around the shock ramp. A semi-analytical theory  was developed to describe SSDA mechanism \cite{Katou2019} which is proved to be consistent with in-situ observations of Earth's bow shock where scattering happens due to whistler waves close to the shock ramp \cite{Amano2020}. 
Further development of this theory \cite{Amano2022} demonstrates that SSDA potentially is able to accelerate electrons very efficiently bridging the injection gap in oblique shocks if proper conditions are met.

Note that in oblique shock simulation electrons sometimes reach the injection energies for the chosen ion-to-electron mass ratio and the shock velocity due to a smaller injection gap ($\varepsilon_{inj,e}/\varepsilon_{th,e} \approx 10^3$ and $(\gamma-1)_{inj,e}\approx35$ for the shock parameters from \cite{Bohdan2022}), however the behaviour of responsible mechanisms in the system with realistic parameters is still not well understood and requires additional studies.

\section{Parallel shocks} \label{sec:par}

\subsection{Shock structure}

At parallel shocks, both ions and electrons are able to overcome the shock barrier and travel back upstream much farther compared to oblique shocks. Figure~\ref{parallel_shock} demonstrates the shock transition region from the longest 2D parallel shocks simulation by \cite{Arbutina2021a,Arbutina2021b}. At the very first stages, the Weibel instability \cite{Weibel,Fried} grows much faster \cite{Crumley2019} than the resonant streaming instability \cite{Zekovic2019}. In the later stages (shown in Fig~\ref{parallel_shock}) the shock is purely mediated by the Alfvenic-like modes which are seeded by the return current via non-resonant streaming Bell instability \cite{Bell2004,Amato2009} \rev{responsible for magnetic field amplification and production of turbulence}. 
The size of the shock (including the foreshock or precursor) is about $d_{sh} \approx 160r_i$, however in hybrid simulations it can be up to $10^3r_i$ \cite{Caprioli2014b}. Unfortunately, the transverse size of the presented 2D simulation does not allow us to capture the full 2D physics  of driven turbulence (e.g. the shock rippling \cite{Wieland2016}) and the shock physics become quasi-1D similar to the 1D simulation done by \cite{Kato2015,Park2015}. Note also that simulation with small ion-to-electron mass ratios can not capture the entire spectrum of driven turbulence. For example the intermediate scale instability can be only captured in simulations with the realistic mass ratio of shocks with moderate Mach numbers $\ma \approx 20$ \cite{Shalaby2021}. 

Therefore, more 2D and 3D simulations, including proper parameters scan for better estimations of the upstream turbulence and its ability to scatter particles, are needed.

\subsection{Electron heating}

\begin{figure}
    \centering
    \includegraphics[width=.99\linewidth]{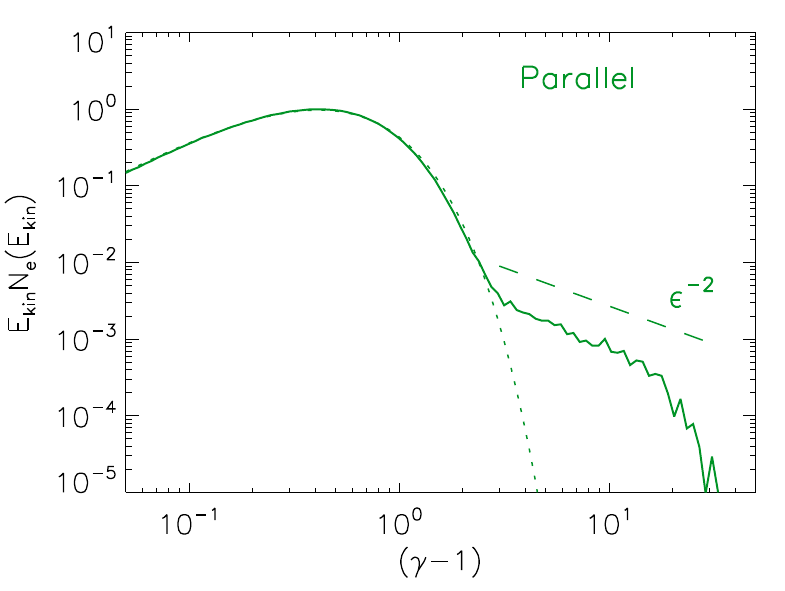}
    \caption{The downstream electron energy distribution in parallel shock ($\thbn=0^\circ$). The dotted lines represent fits of a relativistic Maxwellian to the low-energy part of the spectra. The straight dashed line denotes a power-law $\varepsilon^{-2}$.
    The shock parameters are $\vsh=0.4c$, $\ma=16$, $\mss=35$, $\mi/\me=16$ \cite{Arbutina2021a,Arbutina2021b}.}
    \label{parallel_temp}
\end{figure}

2D  PIC simulations by \cite{Arbutina2021a,Arbutina2021b} suggest that electrons and ions are almost in thermal equilibrium in the downstream ($T_e/T_i\approx (0.5-1)$) of the parallel shock which was also observed in earlier 1D simulations ($T_e/T_i\approx 1$ in \cite{Park2015}, $T_e/T_i\approx 0.5$ in \cite{Kato2015}). However, these results should be treated with some caution. Firstly, 1D simulations are  unable to reproduce certain instabilities. For example, the Weibel instability, which is known to be present in perpendicular shocks, is not excited in 1D simulations \cite{Shimada2000,Hoshino2002} or 2D simulation with an out-of-plane magnetic field configuration \cite{Matsumoto2012,Bohdan2017} though can be triggered only in 2D simulations with in-plane configuration or 3D shock simulation \cite{Matsumoto2017}. Secondly, $T_e/T_i$ may depend on the shock parameters not discussed in \cite{Kato2015,Park2015,Arbutina2021a,Arbutina2021b} and should be studied carefully using 2D and 3D PIC simulations.

\subsection{Electron Acceleration}

The longest 2D (narrow box) parallel shock simulation by \cite{Arbutina2021a,Arbutina2021b} demonstrates the formation of a nonthermal tail both in electron and ion spectra. Indeed, the conditions generated upstream by the shock turbulence in parallel shocks host appropriate conditions for particle scattering and efficient acceleration. The nonthermal tail of the electron energy spectra can be represented by a power-law distribution $\sim \varepsilon^{-2}$ (Fig~\ref{parallel_temp}). The same result was obtained with 1D simulation \cite{Park2015} which suggests that the relevant acceleration physics is similar in 1D and 2D shocks. \cite{Park2015} demonstrated that the upstream cold electrons, after being reflected off the shock because of magnetic mirroring \cite{Ball2001}, remain trapped between the shock front and the upstream waves. At each consequent interaction with the shock, the electron may undergo a new cycle of SDA, which results in strong energy gain. This combination of SDA and scattering on Bell-like waves is also very likely responsible for the formation of the nonthermal electron population in 2D simulations \cite{Arbutina2021a,Arbutina2021b}. However, the applicability of the described mechanisms still should be verified with large 2D and 3D simulations covering all multidimensional effects.

As is the case in oblique shocks, in \cite{Arbutina2021a,Arbutina2021b} electrons reach the injection energy demonstrating that electrons  can be indeed injected into the DSA process. However, the difference between the thermal and injection energies is much smaller ($\varepsilon_{inj,e}/\varepsilon_{th,e} \approx 50$ and $(\gamma-1)_{inj,e}\approx10$) than in SNR shocks. Therefore, we still need to understand how the obtained results can be extrapolated to real systems. Also the electron preacceleration efficiency may depend on the shock velocity, the mass ratio or Mach number and should be clarified in further studies for successful rescaling up to the realistic parameter range.

\section{Summary}

Perpendicular, oblique and parallel shocks are studied to a different degree mostly due to a vast difference in complexity of these systems and amount of computational resources needed to study them. \rev{Also, even if we know the general structure of a shock with predefined upstream conditions, the usage of non-realistic parameters (the shock velocity, mass ratio, etc.) does not permit for a direct application to SNR shocks or in-situ measurements. In many cases (especially oblique and parallel shocks) we still need more PIC simulations to study scaling properties of the shock microphysics for reliable comparison with real shocks and testing of theoretical models.}
All results discussed in this paper are summarized in Table~\ref{tab_results}.

\emph{Perpendicular shocks.} -- We understand all three aspects of the perpendicular shocks physics and can successfully rescale simulations results to compare them with in-situ measurements of planetary bow shocks. The shock is mediated with Buneman and Weibel instabilities. \rev{The latter is responsible for a strong magnetic field amplification inside the shock.} Electrons can be heated via SSA, the shock potential, magnetic reconnection, SFA and adiabatic compression. The downstream electron spectra is represented by the Maxwell distribution with a not very prominent suprathermal tail populated with electrons mostly accelerated via SFA. The contribution of SSA and magnetic reconnection is minor.
Since DSA can not be self-consistently triggered in perpendicular shocks, this case is not so important in astrophysical sense. However, perpendicular shocks share some similarities in structure with oblique shocks and already obtained knowledge can be used to understand better more complex oblique shocks.

\emph{Oblique shocks.} -- Oblique shocks still require thorough studies, however, we have some understanding of the shock structures and electron acceleration mechanisms. In comparison with perpendicular shocks, two additional types of waves (whistlers and electron-acoustic waves) are excited in the foreshock region by the shock reflected electrons. \rev{Whistler waves together with Weibel instability drive strong electromagnetic turbulence responsible for particle heating and acceleration.}
The electron temperature in the shock downstream is likely to be very similar to that in perpendicular shocks, however we still do not know processes responsible for electron heating and lack an explanation for the observed electron temperature. There are at least six processes responsible for particle acceleration in oblique shocks: SSA, SFA, magnetic reconnection, interaction with whistler and electron-acoustic waves and SSDA. The electron downstream spectra is described by Maxwellian with a noticeable suprathermal tail which can be represented by the relatively soft power-law distribution ($\sim \varepsilon^{-2.7}$). These high energy electrons are either produced during interaction with whistler waves or via SSDA. In simulations electrons can potentially reach the injection energy due to much smaller injection gap compared to real shocks. Therefore, additional studies are still needed to understand scaling properties of already known acceleration mechanisms and apply obtained results to real systems.

\begin{table}[]
    \centering
    \caption{Current status of the electron injection problem.}
    \begin{tabular}{cccc}
    \hline
    \hline
     & Structure &  Heating & Acceleration \\
    \hline
    Perpendicular   &    +   & +  & +\\
    Oblique &  $\pm$   &  --  &  $\pm$ \\
    Parallel &  $\pm$  & -- &  $\pm$ \\
    \hline
    \end{tabular}
\begin{tablenotes}
      \small
      \item '$+$' - well studied, '$\pm$' - partially studied, '$-$' - not studied 
    \end{tablenotes}
    \label{tab_results}
\end{table}

\emph{Parallel shocks.} -- Parallel shocks began to be intensively studied only recently, therefore, we already have some understanding of their structure and electron acceleration mechanisms. The shock is mediated by Bell-like waves at its late stage evolution \rev{creating the turbulent precursor where $\delta B/B \sim 1$}. Unfortunately, very high simulation costs restrict transverse size of simulations done so far and at final phases the shock structure is rather one-dimensional. Electrons and ions are in thermal equilibrium, however we still do not have an explanation for that. The electron downstream energy distribution is described by Maxwellian with the prominent suprathermal tail which can be represented by a power-law distribution ($\sim \varepsilon^{-2}$). The most energetic particles are accelerated via a combination of SDA with scattering provided by the foreshock waves. As in oblique shocks, electrons can reach the injection energy in simulations, however the proper extrapolation of simulation results is still needed to derive realistic injection efficiency. The largest problem is that the known physics tends to be 1D (due to the small transverse size of available 2D simulations) and more simulation should be performed to clarify an influence of multidimentinal effects on the electron injection.

\section*{Acknowledgements}

The author thanks Vladimir Zekovi\'c and Bojan Arbutina for sharing data of the parallel shock simulation to reproduce Figures~\ref{parallel_shock} and~\ref{parallel_temp}.

\end{document}